\documentclass[fleqn,usenatbib]{mnras}

\usepackage{newtxtext,newtxmath}

\usepackage[T1]{fontenc}

\DeclareRobustCommand{\VAN}[3]{#2}
\let\VANthebibliography\thebibliography
\def\thebibliography{\DeclareRobustCommand{\VAN}[3]{##3}\VANthebibliography}

\usepackage{graphicx}	
\usepackage{amsmath}	
\usepackage{caption}
\usepackage{subcaption}

\title[Excess Anisotropy of the Radio Background]{Diffuse Sources, Clustering and the Excess Anisotropy of the Radio Synchrotron Background}

\author[F. J. Cowie et al.]{
F. J. Cowie,$^{1,2}$\thanks{E-mail: fraserjcowie@gmail.com (FJC)}
A. R. Offringa,$^{2,3}$
B. K. Gehlot,$^{3}$
J. Singal,$^{4}$
S. Heston,$^{5}$
S. Horiuchi,$^{5,6}$
D. M. Lucero$^{5}$
\\
$^{1}$Department of Physics and Astronomy, The University of Manchester, Oxford Road, Manchester, M13 9PL, UK\\
$^{2}$Netherlands Institute for Radio Astronomy (ASTRON), Oude Hoogeveensedijk 4, 7991 PD Dwingeloo, Netherlands\\
$^{3}$Kapteyn Astronomical Institute, P.O. Box 800, 9700 AV Groningen, Netherlands\\
$^{4}$Physics Department, University of Richmond, 138 UR Drive, Richmond, VA 23173, USA\\
$^{5}$Department of Physics, Virginia Polytechnic and State University, Blacksburg, VA 24061-0435, USA\\
$^{6}$Kavli IPMU (WPI), UTIAS, The University of Tokyo, Kashiwa, Chiba 277-8583, Japan
}

\date{Accepted XXX. Received YYY; in original form ZZZ}

\pubyear{2022}

\begin{document}
\label{firstpage}
\pagerange{\pageref{firstpage}--\pageref{lastpage}}
\maketitle

\begin{abstract}

We present the largest low frequency (120~MHz) arcminute resolution image of the radio synchrotron background (RSB) to date, and its corresponding angular power spectrum of anisotropies (APS) with angular scales ranging from $3^\circ$ to $0.3^\prime$.  We show that the RSB around the North Celestial Pole has a significant excess anisotropy power at all scales over a model of unclustered point sources based on source counts of known source classes.  This anisotropy excess, which does not seem attributable to the diffuse Galactic emission, could be linked to the surface brightness excess of the RSB.  To better understand the information contained within the measured APS, we model the RSB varying the brightness distribution, size, and angular clustering of potential sources. We show that the observed APS could be produced by a population of faint clustered point sources only if the clustering is extreme and the size of the Gaussian clusters is $\lesssim 1'$. We also show that the observed APS could be produced by a population of faint diffuse sources with sizes $\lesssim 1'$, and this is supported by features present in our image. Both of these cases would also cause an associated surface brightness excess. These classes of sources are in a parameter space not well probed by even the deepest radio surveys to date. 

\end{abstract}

\begin{keywords}
radio continuum: general -- diffuse radiation -- radiation mechanisms: non-thermal -- techniques: interferometric -- techniques: image processing
\end{keywords}



\section{Introduction}
\label{sec:intro}

In the last 10 years interest in the radio background has reignited with recent measurements of the radio monopole component. A bright radio background had been measured throughout the late 20th century (e.g., \citealt{costain_1960}, \citealt{haslam_1982}). More recently, combining the surprisingly high monopole component (or surface brightness) measurement of the ARCADE~2 \citep{fixsen_2011} absolutely calibrated stratospheric balloon experiment with lower frequency radio maps which have absolute zero levels, as done in \cite{dowell_2018}, shows a power-law spectrum background of the form:
\begin{equation}
    T_{\text{BGND}}(\nu)=30.4\pm 2.6 \: \text{K} \left(\frac{\nu}{310\:\text{MHz}}\right)^{-2.66\pm 0.04} + T_{\text{CMB}},
	\label{eq:background}
\end{equation}
where $T_{\text{CMB}}$ is a frequency independent contribution from the 2.725~K black-body CMB. The measured spectral index of the background is characteristic of synchrotron radiation and so following convention in the field \citep[e.g.,][]{singal_2018} we refer to this background as the radio synchrotron background (RSB).   This background dominates at frequencies below $\sim 0.5\:\text{GHz}$ and at higher frequencies is below the level of the cosmic microwave background (CMB).  At present, the origin of the radio background is unknown, although several potential explanations have been investigated \citep[e.g.,][]{singal_2018,singal_2022}.  

One potential cause of the radio background is extragalactic radio sources.  However recent works based on deep radio source counts have shown that known classes of extragalactic radio sources can only contribute around one-fifth of the measured radio background brightness (e.g., \citealt{condon_2012}, \citealt{vernstrom_2014}, \citealt{hardcastle_2020}). To attribute the measured radio background to point sources would require a new, so far unobserved population of extremely numerous and faint sources. These sources would likely be of a new physical origin as they would have to have a density at least an order of magnitude greater than galaxies in the Hubble Ultra Deep Field \citep{condon_2012}. Alternative explanations for the excess background include classes of diffuse extragalactic sources such as dark matter annihilation and decays (e.g., \citealt{fornengo_2011}, \citealt{hooper_2012}) or cluster mergers (e.g., \citealt{fang_2016}). Another possible explanation is that the excess could be caused by a large, bright, approximately spherical synchrotron halo surrounding the Milky Way galaxy (e.g., \citealt{subrahmanyan_2013}). However, this would make the Milky Way unique among nearby spiral galaxies \citep{singal_2015} and would drastically change our theory and understanding of the high-latitude Galactic magnetic field \citep{singal_2010}.

While several measurements of the radio background surface brightness have been made, few experiments have explored its anisotropy. Anisotropy studies and measurement have greatly contributed to constraining source populations responsible for other cosmic backgrounds, such as the infrared (e.g., \citealt{ade_2011}, \citealt{george_2015}) and gamma-ray (e.g., \citealt{broderick_2014}) backgrounds. While not a typical cosmic background produced by a discrete population of sources, large scale, precise measurements of the CMB anisotropy have been instrumental.

The earliest measurements of the anisotropy of the RSB come from searches for CMB anisotropies at low frequencies. These measurements are all at frequencies where the RSB surface brightness is more than an order of magnitude lower than the CMB and are decades old. Additionally, these measurements are confusion noise limited on the anisotropy at certain discrete scales, and have small fields of view in comparison to the current work. These anisotropy measurements are made with the VLA at 8.4~GHz \citep{partridge_1997} and at 4.9~GHz \citep{fomalont_1988}, and the Australia Compact Telescope Array at 8.7~GHz \citep{subrahmanyan_2000}. These results and their respective constraints on the RSB anisotropy are summarised in \cite{holder_2014}. More recent and more complete measurements of the RSB anisotropy have been made with the Giant Metrewave Radio Telescope, presented in \cite{choudhuri_2020}. However, this experiment was primarily focused on the measurement of the power spectrum for 21-cm studies of the epoch of reionization, and was more limited in angular range. The question on the origin of the anisotropies within the RSB was also not addressed. Similarly, \cite{bernardi_2009} and \cite{ghosh_2012} measured the anisotropy power spectrum of the RSB also at 150~MHz in order to characterise the foregrounds for epoch of reionization experiments, and \cite{iacobelli_2013} measured the anisotropy power spectrum to study interstellar turbulence. On larger angular scales, with only a small overlap with what is presented here, determinations of the anisotropy power spectrum where large scale Galactic diffuse synchrotron emission dominates have been made \citep{gehlot_2022}. 

\cite{offringa_2022} presented the first targeted measurement of the anisotropy power spectrum of the RSB, which was made at 140~MHz. That result showed an unexplained excess anisotropy power in the RSB. This work goes further and presents measurements at a lower frequency and over a larger range of angular scales. We perform more robust tests for systematic uncertainties, including full pipeline simulations for radio background models. For the first time, we discuss the implications of the measured anisotropy of the RSB using extensive modelling of the radio background as a reference.

In this work we present the largest arcminute resolution image of the radio background to date and its corresponding angular power spectrum of measured anisotropies (APS) with angular scales ranging from $3^\circ$ to $0.3^\prime$. These measurements are based on dedicated Low-Frequency Array \citep[LOFAR --][] {lofar} observations at 120~MHz of seven 64 $\text{deg}^2$ fields. \S\ref{sec:obs} describes the observations, their data reduction, and a demonstration of their flux calibration. \S\ref{sec:modelling} outlines the methods used to simulate images of the radio sky. \S\ref{sec:results} presents the measured angular power spectra from the observations, alongside spectra from simulated images where the sources are distributed according to various combinations of source count and clustering models. \S\ref{sec:disscussion} discusses the implications of the observations for constraining the possible origins of the RSB.

\section{Observations and Data Reduction}
\label{sec:obs}

\begin{table}
 \caption{Summary of the observational details valid for all pointings observed.}
 \label{tab:obs_details}
 \begin{tabular*}{\columnwidth}{@{}l@{\hspace*{48.5pt}}l@{}}
 \hline
 Observing project & LC9\_007 \\
 Observation start time (UTC) & 2018/03/05 17:44:37.0\\
 Observation end time (UTC) & 2018/03/06 05:42:08.8 \\
 Duration & 43051.8 s ($\sim$ 11.96 h) \\
 Frequency range & 115.76-127.35 MHz \\
 Frequency resolution (after averaging) & 61.035 kHz \\
 Sub-band width & 1.83 MHz \\
 Bandwidth & 11.6 MHz \\
 Central frequency & 120.6 MHz \\
 Number of pointings & 7 \\
 Field of view of single pointing & $\sim 5.6^{\circ}$ \\
 \hline
 \end{tabular*}
\end{table}

\begin{table}
 \caption{Summary of the different fields.}
 \label{tab:field_details}
 \begin{tabular*}{\columnwidth}{@{}l@{\hspace*{12.1pt}}l@{\hspace*{12.1pt}}l@{}}
 \hline
 Field name & Field pointing & Field noise, $\sigma$ (mJy $\text{beam}^{-1}$)) \\
 NCP field A & $00h00m00s$, $+90^{\circ}00^{\prime}00^{\prime\prime}$ & 1.08 \\
 NCP field B & $02h00m00s$, $+86^{\circ}00^{\prime}00^{\prime\prime}$ & 1.3 \\
 NCP field C & $06h00m00s$, $+86^{\circ}00^{\prime}00^{\prime\prime}$ & 1.06 \\
 NCP field D & $10h00m00s$, $+86^{\circ}00^{\prime}00^{\prime\prime}$ & 1.12 \\
 NCP field E & $14h00m00s$, $+86^{\circ}00^{\prime}00^{\prime\prime}$ & 0.96 \\
 NCP field F & $18h00m00s$, $+86^{\circ}00^{\prime}00^{\prime\prime}$ & 1.11 \\
 NCP field G & $22h00m00s$, $+86^{\circ}00^{\prime}00^{\prime\prime}$ & 1.2 \\
 \hline
 \end{tabular*}
\end{table}

The data used in this analysis was from approximately 12 hours of observing with LOFAR using the high-band antenna (HBA) dual mode, with multi-beaming and using only Dutch stations. Observations were in the frequency band from 114 to 126 MHz on the night of March 5, 2018. Multi-beaming allows for 7 adjacent fields to be observed simultaneously. The central field (field A) was chosen to be centred on the north celestial pole. The 6 flanking fields were at declination of $+86^{\circ}00^{\prime}00^{\prime\prime}$ and equally spaced in right ascension. One of these fields has been previously analysed for epoch of reionization science by \cite{gan_2022}. The first of these flanking fields was at a right ascension of $2h00m00s$. The north celestial pole was chosen for this measurement due to the relatively low Galactic component to the background, as can be seen from the 408 MHz all sky map by \cite{haslam_1982}; as well as overlap with other observations which make future cross-correlation analyses possible; and the abundance of data (over 600 hours) due to an overlap with the LOFAR epoch of reionization field. Alongside the target fields the flux calibrator 3C\,147 was observed. The observational parameters are summarised in Table~\ref{tab:obs_details}. The different pointings are summarised in Table~\ref{tab:field_details}.

The raw data is processed using the LOFAR Initial Calibration ({\sc{linc}}) pipeline \citep{van_weeren_2016,williams_2016} and only direction-independent calibration is performed. Direction-dependent calibration is not necessary, because we only analyse scales $\geq 30''$, for which the ionospheric effects are negligible. Direction-dependent calibration may also introduce systematic effects \citep{sardarabadi_2019, mevius_2022} which are avoided this way. The {\sc{linc}} pipeline makes use of many software packages including the Default Pre-Processing Pipeline ({\sc{dp3}}; \cite{van_diepen_2018}), LOFAR SolutionTool ({\sc{losoto}}; \cite{de_gasperin_2019}) and {\sc{aoflagger}} \citep{offringa_2012}. This pipeline has been used for previous measurements of the anisotropy power spectrum of the radio background with LOFAR, and the results were found to be similar to manual calibration pipeline \citep{offringa_2022}. All baselines greater than $21$ km were flagged using {\sc{dp3}}. Long baselines add negligible sensitivity when imaging diffuse structure and begin to resolve unwanted ionospheric effects.

Each of the 7 fields were then deconvolved using a {\sc{wsclean}} multi-frequency multi-scale deconvolution \citep{offringa_2012} with auto-masking and uniform weighting. During the cleaning the dirty beam had a size of approximately $20'' \times 20''$ and the fields were imaged with a field of view of approximately 8 degrees. All subsequent imaging was also done using {\sc{wsclean}}. The auto-masking was used to ensure all sources above $\geq 5 \sigma$ were modelled to a $1 \sigma$ level. In this case $\sigma$ was approximately {1 mJy} in all fields, see Table~\ref{tab:field_details} for exact values. In \citet{offringa_2022}, a multi-scale clean was not used so as to avoid subtraction of potential diffuse signal. However, due to the larger field of view of this measurement, a significant number of resolved sources were present, so in order to model these sources as accurately as possible a multi-scale clean is needed. The sky models created from the cleaning were inspected to ensure a large scale diffuse background component was not being subtracted alongside the resolved sources as a consequence of multiscale cleaning. The prediction step in {\sc{wsclean}}, using the default w-stacking algorithm \citep{offringa_2014}, was then used to model visibilities based on the sky model generated during the cleaning. These model visibilities were then subtracted from the data, leaving only sources $\leq 5 \sigma$, any diffuse background, and noise. Each of the 7 fields were then imaged using natural weighting of the gridded visibilities, in order to obtain optimal sensitivity of the power spectrum calculated in Section \ref{sec:results}. Additionally, due to accentuation of diffuse structure in natural weighting, the images also allowed for a first qualitative assessment of the background.

\begin{figure}
	\includegraphics[width=\columnwidth]{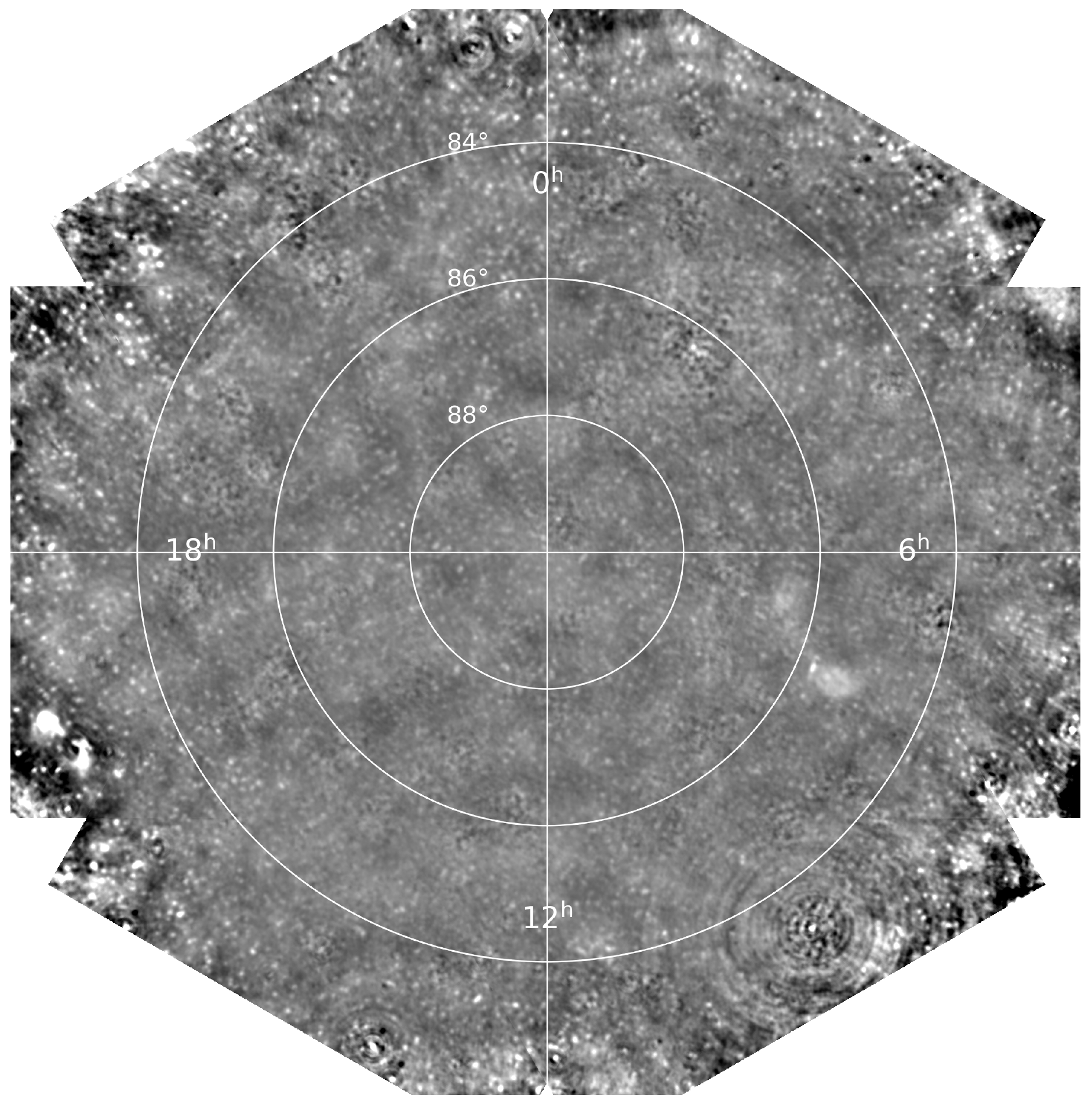}
    \caption{The radio sky around the NCP at 120 MHz as seen by LOFAR. The image has a width and height of approximately 16 degrees. The image is naturally weighted to accentuate the diffuse structure and the synthesised beam has a size of approximately $3.3^\prime \times 3.3^\prime$. Towards the top right the residual emission from the approximately 40 Jy source 3C\,61.1 can be seen. In the middle right of the combined field a currently unidentified patch of highly diffuse emission is present. All channels are collapsed during imaging.}
    \label{fig:mosaic_background}
\end{figure}

The 7 fields were then mosaicked together using a resampling method. This was done using the Python package {\sc{reproject}} \citep{reproject} with bilinear interpolation. The mosaicking was done taking into account the primary beam response of LOFAR for each observed field. These were calculated using the software package {\sc{EveryBeam}}\footnote{https://git.astron.nl/RD/EveryBeam}. The mosaicking was done both for the background source subtracted images and the cleaned component images. The final result of the data reduction was a naturally weighted, flux calibrated, $200~\text{deg}^2$ image of the diffuse radio background at the NCP. The synthesised beam has a size of approximately $3.3^\prime \times 3.3^\prime$. The image is shown in Fig.~\ref{fig:mosaic_background}.

\begin{figure}
	\includegraphics[width=\columnwidth]{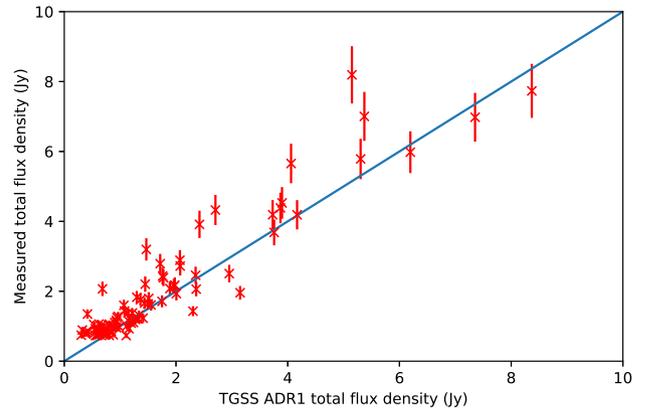}
    \caption{The measured total flux density from the cleaned mosaic against the total flux density from the TGSS ADR1 source catalogue \protect\citep{intema_2017}. The total flux densities from the source catalogue have been adjusted to represent the expected values at 120~MHz. The error bars represent a 10\% error on the measured total flux density. The brightest source, 3C\,61.1, is not included for clarity, but lies within 10\% of the source catalogue value.}
    \label{fig:calibration}
\end{figure}

\begin{figure}
	\includegraphics[width=\columnwidth]{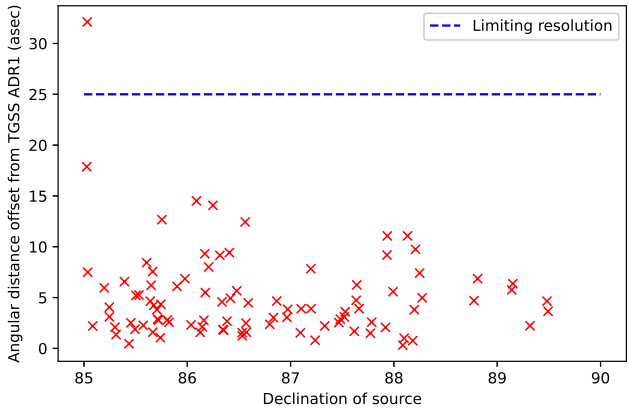}
    \caption{The angular distance offset of cross matched sources in the cleaned mosaic from their matches in the TGSS ADR1 source catalogue, as a function of declination of the sources. The dotted blue line represents the limiting resolution, in this case the $25^{\prime\prime}$ resolution of the TGSS ADR1 source catalogue.}
    \label{fig:offsets}
\end{figure}

In order to check that the observations were flux calibrated correctly during the data reduction stage, the mosaic of clean components was used as input for the software PYthon Blob Detector and Source Finder ({\sc{pybdsf}}; \citealt{pybdsf}). The 100 brightest sources at a declination greater than $+85^{\circ}00^{\prime}00^{\prime\prime}$ were all found to have counterparts with the source catalogue released as part of the TGSS first alternative data release (ADR1) \citep{intema_2017}. The declination cut was used to evaluate the flux calibration for the area of the image corresponding to the area of interest for producing the anisotropy power spectrum. As the TGSS observations are at a frequency of 150~MHz the total flux density from the source catalogue was adjusted to the expected at 120~MHz using a spectral index of -0.7, which is characteristic of the RSB \citep{offringa_2022}. Figure~\ref{fig:calibration} shows the measured total flux density from the cleaned mosaic against the total flux density from the source catalogue, where the error bars represent a 10\% error on the measured total flux density. This shows that our observations are well calibrated in flux density as most of the 100 brightest sources lie within 10\% of the source catalogue value. Figure~\ref{fig:offsets} shows the angular distance offsets of sources identified in the cleaned mosaic from their cross matched sources in the source catalogue. The majority of sources have angular distance offsets less than the limiting resolution of the TGSS ADR1 observations showing that our observations are well calibrated.

\section{Modelling the radio synchrotron background}
\label{sec:modelling}

In order to draw conclusions from the observed anisotropy power of the RSB comparisons between observations and theory must be made. We model several different source populations with differing spatial clustering, flux density distributions and angular size to investigate the effect of these variables on the anisotropy power spectrum. The general procedure for modelling a population of sources was as follows. A differential source count form was chosen and using the method of inverse transform sampling a list of sources with flux density values consistent with the flux density distribution was generated. The number of sources generated in a given solid angle was such that it was consistent with the differential source count. In many cases the flux distribution of sources used is based off observations at a different frequency. In order to account for this, each source is also assigned a spectral index by drawing samples from a normal distribution centred on the spectral index of the RSB, taken to be $-0.6$ to be consistent with typical astrophysical synchrotron radiation \citep{rybicki_1986}, with a standard deviation of $0.1$. The flux density of the sources could then be adjusted to the relevant central frequency using the generated spectral indices. Each source is then assigned a position on the sky. These positions are either generated so that sources are uniformly distributed on the sky, or so that they are clustered in some way. A specific angular size is chosen for the sources, or they are specified to be point sources. This is to allow for different hypothetical source populations to be simulated, both point sources and resolved sources are postulated in the literature as solutions to the observed monopole component excess of the RSB (see Section~\ref{sec:intro}). If the sources are of a specific size, then they are rendered as smooth Gaussians on the sky with the angular size referring to their full width half maximum in flux. This catalogue of sources is then placed onto a grid of pixels using sinc interpolation, producing a model image that is an accurately downsampled version of the continuous sky model. The relevant Fourier modes of the pixelated image match the Fourier modes of the continuous sky model up to the machine precision. Unless otherwise stated, the grid size of the simulations was 2880x2880 and the image size was $16^\circ$x$16^\circ$. The resulting model image is then used as an input in the {\sc{wsclean}} prediction step in order to generate the visibilities the LOFAR would observe for the given model of sources. The visibilities are then imaged in the same way as an individual field as described in Section \ref{sec:obs}. Because this concerns only a single beam, no mosaicking is necessary. This resulted in model images of the radio sky.

The two differential source count models used in simulating the radio sky were the semi-empirical model of observed source counts presented in \cite{franzen_2016}, and a hypothetical model presented in \cite{condon_2012} which matches the radio background monopole component excess. The latter is achieved through the existence of a very large population of faint, so far undetected sources. These models will be referred to here as the Franzen and Condon models respectively. The Franzen model has a source density as a function of flux ($n(S)$) of the form:
\begin{equation}
\label{franzen}
    n(S) = \frac{dN}{dS} = k \left( \frac{S}{\text{Jy}} \right)^{-\gamma} \text{Jy}^{-1} \: \text{sr}^{-1} \: \text{for 0.1 mJy} \leq S \leq 400 \: \text{mJy}, 
\end{equation}
where $k=6998$ and $\gamma=1.54$. This is a source count at 154~MHz and corresponds to model A in Table 2 of \cite{franzen_2016}. The Condon model has a source density of the form:
\begin{equation}
\label{condon}
    n(S) = \frac{A}{S^2} \text{exp}\left(-4 \ln{2} \frac{(\log({S}) - \log(S_{\text{pk}}))^2}{\phi^2}\right) \text{Jy}^{-1} \: \text{sr}^{-1},
\end{equation}
where $\phi=0.2$, $S_{\text{pk}}=39 \: \text{nJy}$, $\log({A(\text{Jy} \: \text{sr}^{-1}))} = 4.67$ and $S$ has units of Jy. This is a source count at 1.4~GHz and corresponds to the model with the fewest sources sufficient to make the surface brightness of the RSB presented in \cite{condon_2012}.

Clustering of source populations can change the anisotropy power on different scales. In \cite{offringa_2022} sinusoidal clustering on different scales was explored and found to have varying effects on the anisotropy power spectrum. In this work, we focus on a more general and perhaps more physically realistic clustering scheme, which we will refer to as Gaussian clustering. In this method, sources are assigned to clusters which have a 2D Gaussian density profile, and the clusters themselves are uniformly distributed across the sky. Free parameters of the clustering are the number of clusters within a given solid angle and the angular size of the Gaussian cluster, effectively the standard deviation of the 2D Gaussian. The number of sources in each cluster is implied from the number of clusters within a given solid angle, because the total number of sources is kept constant. A mathematically rigorous normal distribution on a sphere is known as a Kent distribution and is not trivial (see \citealt{kent_1982}), so due to computational limits true Kent distributions are instead approximated. 

A Gaussian cluster is first created and populated at a declination of $+90^{\circ}00^{\prime}00^{\prime\prime}$. This allows for the cluster to be populated by drawing positions of sources belonging to the cluster from a uniform random distribution in right ascension, and a normal distribution in declination, with the desired size of the cluster as the standard deviation. The whole cluster is then rotated to a random direction on the sky such that the clusters themselves are uniformly distributed across the sky. This process is then repeated to generate each cluster. This approximation is valid for cases where the cluster size is less than 1 degree. Having a cluster's size much larger than this invalidates the step where the cluster was populated using a normal distribution in declination, as this will no longer be a good approximation of the Kent distribution.

\section{Results}
\label{sec:results}

\subsection{Observed angular power spectra}
\label{sec:observed_results}

\begin{figure}
	\includegraphics[width=\columnwidth]{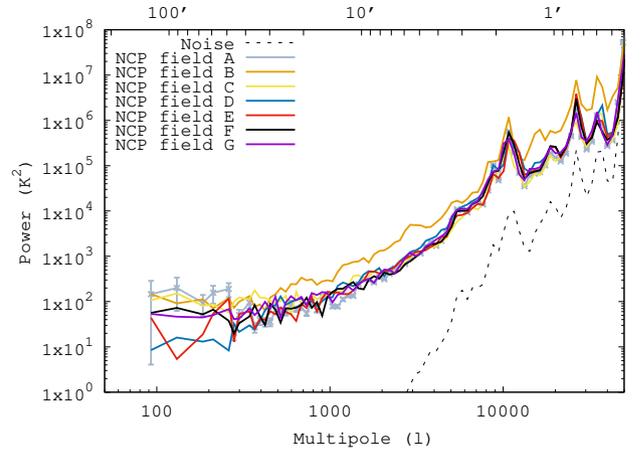}
    \caption{The measured APS of the radio sky for the different fields around the NCP at a central frequency of 120 MHz. Error bars  are only shown for one field for clarity and represent the $2\sigma$ errors due to cosmic variance. The top axis denotes the angular scale equivalent to the multipole on the bottom axis. The dotted line is the APS due to the noise present in field A. All fields have similar anisotropy power apart from field B. This is likely because field B has the bright source 3C\,61.1 close to its centre. }
    \label{fig:all_ps}
\end{figure}

The APS of the resulting naturally weighted image from observations or modelling is generated using the power spectrum pipeline described in \cite{offringa_2022} which is originally based on the power spectrum pipeline described by \cite{offringa_2019}. The APS of the central $3.8^\circ$x$3.8^\circ$ patch of the 7 different fields are shown in Fig.~\ref{fig:all_ps}. This is computed for the fields without first correcting for the primary beam of the instrument. For each field, all sources $\geq 5 \sigma$ are subtracted to a $1 \sigma$ level, as described in Section~\ref{sec:obs}. The APS of field B becomes power law like at lower $\ell$ and shows an excess over the other fields in the power law region. This is likely because field B has the bright source 3C\,61.1 close to its centre. The excess anisotropy power likely arises due to residuals or artefacts from this bright source that are not subtracted completely. The contribution to the APS of the fields from the noise is estimated by making two naturally weighted images using the pre-cleaning visibilities from odd time steps and even time steps respectively. Subtracting these two images from each other resulted in a noise image where the noise is representative for an observation of half the integration time. Therefore, the image is divided by $\sqrt{2}$ to obtain a true noise image for our observations. This process is only done for field A, as while there may be variation in the noise due to the differing elevation of the fields, this is negligible. The noise image is then processed using the APS pipeline and the resulting APS is shown in Fig.~\ref{fig:all_ps}. The error bars shown on all APS presented throughout are $2\sigma$ errors due to cosmic variance and do not include systematic effects unless otherwise stated. Errors due to cosmic variance are those of sample variance because at each value of $\ell$ a finite number of angular modes are sampled to calculate the APS, due to the finite field of view used. The physical scale size of an angular mode corresponding to a certain $\ell$ is well approximated by $\theta_{\ell} = \frac{180}{\ell}$. Then, the number of independent modes sampled by a field of view with angular scale $\Theta$, for a certain $\ell$, is $N_{\ell} = \left(\frac{\Theta}{\theta_{\ell}}\right)^2$. The $2\sigma$ fractional error due to cosmic variance on the APS is then given by $\frac{2}{\sqrt{N_{\ell}}}$, assuming Poisson statistics. The APS of the noise image shows that the noise power is insignificant compared to the observed power except on scales $\ell>10000$, which are not our focus. However, the noise APS shows the presence of systematic peaks. These peaks correspond to scales where the instrument has low sensitivity, arising from poor $uv$-coverage of the instrument at certain scales. These align with peaks in the actual data, and this indicates some form of systematic causes an apparently multiplicative effect when the sensitivity of the instrument is low. This effect was seen previously in \cite{offringa_2022} and will henceforth be referred to as instrumental power spectrum systematics to distinguish them from other systematic effects such as calibration errors. These peaks are present to some degree in all modelled APS shown in Section~\ref{sec:modelled_results}, however are exacerbated in the full pipeline simulation which includes source subtraction, as seen in Section~\ref{sec:unsubtracted}. This suggests the main origin of these instrumental power spectrum systematics occurs during the subtraction process.

\begin{figure}
	\includegraphics[width=\columnwidth]{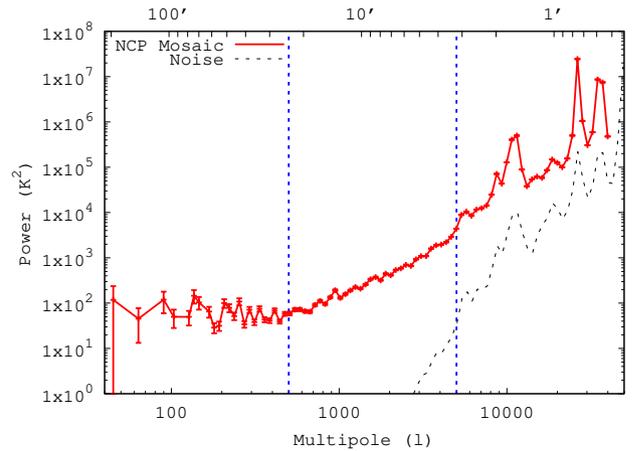}
    \caption{The measured APS of the radio sky around the NCP at a central frequency of 120 MHz. The solid curve shown is the APS of the mosaicked image and the error bars represent the $2\sigma$ errors due to cosmic variance. The dotted line is the APS due to the noise present in field A, and can be assumed to be a representative upper limit of the noise contribution to the mosaicked APS. The dotted blue vertical lines split the APS into 3 sections which exhibit different characteristic behaviour as discussed in \S \ref{sec:observed_results}.}
    \label{fig:mosaic_ps}
\end{figure}

The APS of the central $8^\circ$x$8^\circ$ patch of the mosaicked image is shown in Fig.~\ref{fig:mosaic_ps}. Only the central part of the mosaicked image is used as this is where the sensitivity is greatest. The noise APS of field zero is also shown, although it is noted that the actual noise contribution to the mosaicked image APS is likely less than this, due to overlapping of fields during the mosaicking.

The APS of the mosaicked image can be best interpreted by splitting it into 3 sections, as depicted in Figure~\ref{fig:mosaic_ps}. Firstly, the flat part of the spectrum from $50\leq \ell \leq 700$  has a shape consistent with what is expected from diffuse Galactic emission \citep{gehlot_2022}. Additionally, the anisotropy power observed is consistent with independent AARTFAAC measurements of the NCP done by \cite{gehlot_2022}. Secondly, the power law part of the spectrum for $700 \leq \ell \leq 4000$ is indicative of a region where unclustered point sources dominate the anisotropy power \citep{tegmark_1996}, well fit by a power law with index $\beta = 2.17 \pm 0.08$. However, as discussed in Sections~\ref{sec:modelled_results} and \ref{sec:unsubtracted} this shape could also be produced by other classes of source populations. Finally, at the smallest scales, $\ell \geq 4000$, the anisotropy power is dominated by instrumental power spectrum systematics. These peaks make it infeasible to conclude anything about the measured anisotropy power for $\ell > 4000$.

\subsection{Modelled angular power spectra}
\label{sec:modelled_results}

\begin{figure*}
\begin{subfigure}[t]{0.49\textwidth}
	\includegraphics[width=\columnwidth]{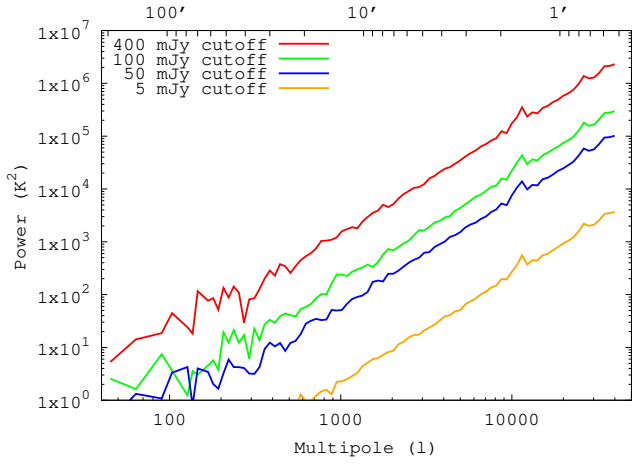}
    \caption{APS of unclustered Franzen models with different brightness distribution cutoffs in the model. The models are over a $8^\circ$x$8^\circ$ patch of sky.}
    \label{fig:cutoff_ps}
\end{subfigure}
\hfill
\begin{subfigure}[t]{0.49\textwidth}
	\includegraphics[width=\columnwidth]{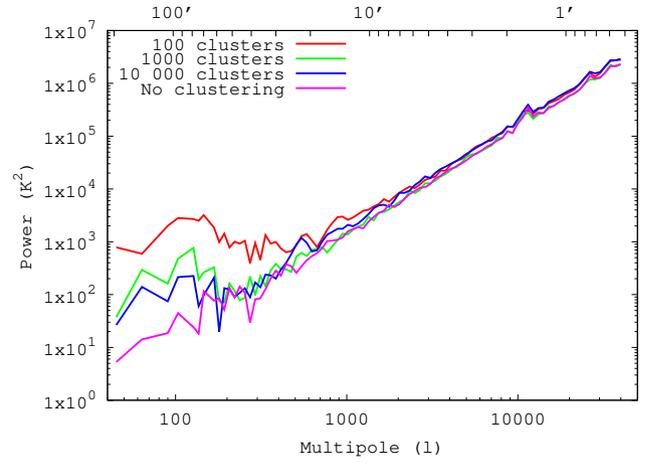}
    \caption{APS of clustered Franzen models with different numbers of Gaussian clusters in the model. The sizes of the clusters are fixed at 1 degree. The unclustered Franzen model APS is shown for reference. The models are over a $8^\circ$x$8^\circ$ patch of sky.}
    \label{fig:ncluster_ps}
\end{subfigure}
\vskip\baselineskip
\begin{subfigure}[t]{0.49\textwidth}
	\includegraphics[width=\columnwidth]{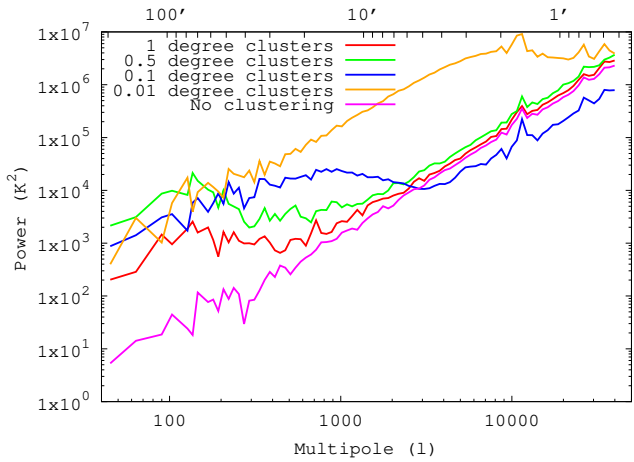}
    \caption{APS of clustered Franzen models with different sizes of Gaussian cluster in the model. The number of clusters was fixed at 100. The unclustered Franzen model APS is shown for reference. The models are over a $8^\circ$x$8^\circ$ patch of sky.}
    \label{fig:clustersize_ps}
\end{subfigure}
\hfill
\begin{subfigure}[t]{0.49\textwidth}
	\includegraphics[width=\columnwidth]{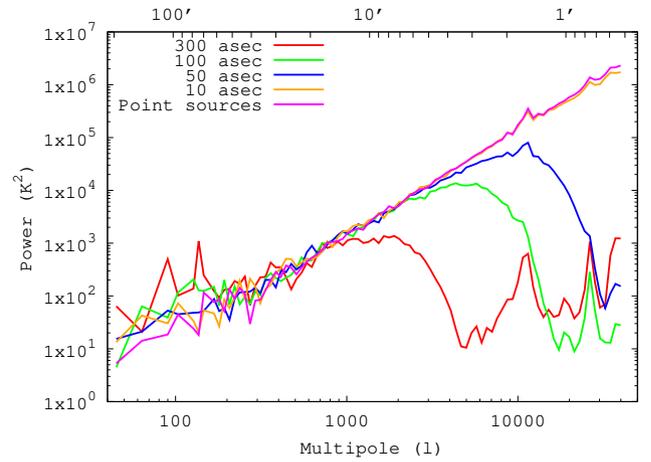}
    \caption{APS of unclustered Franzen models with sources modelled as diffuse Gaussians of different sizes. The unclustered Franzen model with all sources rendered as point sources is shown for reference. The models are over a $8^\circ$x$8^\circ$ patch of sky.}
    \label{fig:sourcesize_ps}
\end{subfigure}
\caption{APS of Franzen models with different clustering, brightness distribution and source size properties. Unless otherwise stated all sources were rendered as point sources and there was no cutoff in the brightness distribution of sources. In all cases the images used to generate the APS are generated from simulated visibilities of the LOFAR instrument. No cleaning is performed on the images.}
\end{figure*}

Several different models are used to create images of the radio sky at 120 MHz in the same format as the observations, following the procedure described in Section \ref{sec:modelling}. APS for these model images are then created by using the same pipeline and the effect on the APS of changing different variables is investigated. Firstly, the effect of different flux cutoffs on the unclustered point source Franzen model is investigated. Figure~\ref{fig:cutoff_ps} shows the APS produced by the Franzen model with sources present up to 400~mJy (no flux density cutoff), 100~mJy, 50~mJy and 5~mJy. Despite the lower flux density sources being orders of magnitude more numerous in the Franzen model, the brightest sources dominate the anisotropy power. The formula for the multipole moment, $C_{\ell}$, for an unclustered population of point sources \citep{tegmark_1996} supports this:
\begin{equation}
\label{unclustered_power}
    C_{\ell} (\nu) = \int_0^{S_{\nu, \text{cut}}} S_{\nu}^{2} \frac{dN}{dS_{\nu}} dS_{\nu},
\end{equation}
where $S_{\nu}$ is the flux density, $\frac{dN}{dS_{\nu}}$ is the differential source count per steradian, and $S_{\nu, \text{cut}}$ is the flux density limit to which sources are removed by subtraction. After converting $C_{\ell}$ from flux density to temperature units (see \citealt{tegmark_1996}), $C_{\ell}$ is related to the plotted anisotropy power, $(\Delta T)_{\ell}^2$, by:
\begin{equation}
\label{normalization_power}
    (\Delta T)_{\ell}^2 = \frac{\ell(\ell + 1)}{2 \pi} C_{\ell}. 
\end{equation}
The shape of the APS is therefore also consistent with what is expected for unclustered point sources \citep{tegmark_1996}. Small peaks due to instrument power spectrum systematics are observed at large $\ell$, and at small $\ell$ some cosmic variance is observed.

We also simulate the APS of the Franzen model with clustered sources using the Gaussian clustering described in Section~\ref{sec:modelling}. The free parameters for the clustering are the angular size of the clusters on the sky, and the number of total clusters within a given solid angle. In this case, the solid angle is the area of sky between a declination of $+75^{\circ}00^{\prime}00^{\prime\prime}$ and $+90^{\circ}00^{\prime}00^{\prime\prime}$. The size of the clusters is fixed at 1 degree and the number of clusters varied. The sources are all generated as point sources and no flux cutoff was used in the Franzen model. Fig.~\ref{fig:ncluster_ps} shows the APS produced by this model with the number of clusters set to 100, 1000, and 10000. The unclustered Franzen model APS is shown for reference. From the APS it is seen that the clustering produces an excess anisotropy power on scales roughly greater than the cluster size. As the number of clusters increases, the number of sources per cluster decreases and the APS approaches that of the unclustered case. Indeed, the unclustered case is equivalent with having one source per cluster, as the clusters themselves are uniformly distributed on the sky. 

In Fig.~\ref{fig:clustersize_ps}, the number of clusters is fixed at 100 while the angular size of the clusters is set to 1, 0.5, 0.1 and 0.01 degrees. The unclustered Franzen model APS is shown for reference. The APS show that clustering produces an excess anisotropy power on scales larger than the size of the cluster. On scales much larger than the size of the clusters, the shape (power law) of the APS converges to an unclustered point source APS, however, the magnitude of the anisotropy power is greater than if the sources were unclustered. This can be explained by the fact that the large scales no longer resolve individual point sources inside a cluster, and therefore the cluster would act as one strong point source on these scales, with a brightness formed from the coherent sum of the sources in the cluster. When sources are resolved, they add incoherently in the anisotropy power. In other words, a single bright point source causes a higher anisotropy power compared to a collection of point sources that add up to the same brightness. This effect is expected from equation~\eqref{unclustered_power} and shown in Figure~\ref{fig:cutoff_ps}, as the brightest sources dominate over the more numerous but fainter sources. On intermediate scales the APS has a flat shape with an anisotropy excess over the unclustered case.

Up to this point only point sources have been considered. We next investigated the effect of having diffuse Gaussian sources of different sizes.  The unclustered Franzen model with no flux cutoff is used and the angular size of the diffuse Gaussian sources is varied. The diffuse sources have the same total flux density as a corresponding point source but a lower surface brightness and peak flux density. All sources were chosen to be a single size for simplicity, however, in principle a distribution of sizes is more realistic. Figure~\ref{fig:sourcesize_ps} shows the APS produced by this model with the source size set to 300, 100, 50, and 10 arcseconds. The APS of the unclustered Franzen model with only point sources is shown for reference. On scales larger than the sources the shape converges to the unclustered point source APS. On scales smaller than the source size the anisotropy power rapidly decays. This is a reflection of the fact that the modelled diffuse Gaussian sources are smooth on scales smaller than the source size. This is one key difference between having a single diffuse Gaussian source and a Gaussian cluster of faint point sources. The cluster will show spatial fluctuations and have non-negligible anisotropy power on scales smaller than the cluster size, as shown in Fig.~\ref{fig:clustersize_ps}.

\begin{figure}
	\includegraphics[width=\columnwidth]{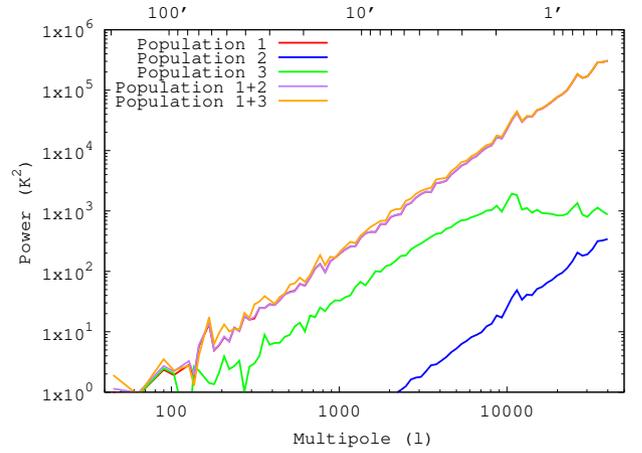}
    \caption{APS of Franzen models with cuts in the brightness distribution such that there is a bright population of sources, population 1, and a faint population of sources, population 2. Population 2 is then clustered using 1000 Gaussian clusters of size 0.01 degrees to give population 3. The APS resulting from the presence of different combinations of these populations are presented. The APS from population 1 is present but hidden by the APS of population 1+2. The images used to generate the APS are generated from simulated visibilities of the LOFAR instrument. No cleaning is performed on the images.}
    \label{fig:twopop_ps}
\end{figure}

To investigate the result of two distinct populations of sources, we perform a hybrid simulation: the Franzen model with upper and lower cuts at 100 and 10~mJy respectively, gives population 1. The Franzen model again is used with an upper cut at 1~mJy to give population 2, representing a population of fainter sources. The sources in population 2 were clustered using Gaussian like clustering to give population 3. This had the number of clusters parameter set to 1000 and the size of the clusters as 0.01 degrees. All sources were simulated as point sources. Figure~\ref{fig:twopop_ps} shows the APS produced by several models with these populations combined and on their own. Adding a faint unclustered population of sources to an existing much brighter population of unclustered sources has a negligible effect on the APS, demonstrated by the fact that the APS for population 1 and populations 1+2 lie on top of each other. However, adding a clustered faint population of sources has an effect on the APS, albeit a small one compared to the relative initial effect of the clustering on increasing the APS of the faint population. 

This demonstrates that one possible way for a very faint population of sources to create an excess anisotropy power over a population of bright sources is for them to be clustered. However, the clustering must be strong for this to occur. Furthermore, clustering causes a power law shaped or multiplicative excess only on scales larger than the size of the cluster as shown in Fig.~\ref{fig:clustersize_ps}. This would mean that any excess anisotropy power decrease on scales smaller than the cluster.

\begin{figure}
	\includegraphics[width=\columnwidth]{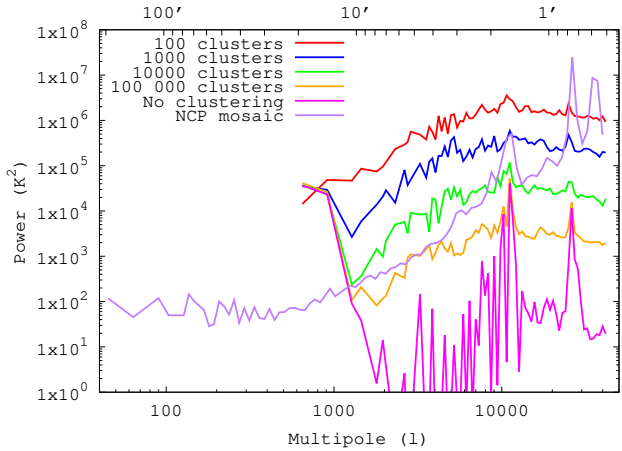}
    \caption{APS of clustered Condon models with different numbers of Gaussian clusters in the model. The sizes of the clusters are fixed at 30 arcseconds. The unclustered Condon model APS is shown for reference. The models are over a $0.75^\circ$x$0.75^\circ$ patch of sky and the grid size is 360x360. The APS of the observed NCP mosaic is shown for reference.}
    \label{fig:condon_ps}
\end{figure}

Finally, the Condon model with and without clustering is explored. The Condon model is used with no brightness upper limits and with all sources rendered as point sources. Figure~\ref{fig:condon_ps} shows the APS of clustered Condon models where the size of the clusters fixed at 30 arcseconds and with the number of clusters set to 100, 1000, 10 000 and 100 000. The observed and unclustered Condon model APS are shown for reference. The unclustered Condon model APS shows high levels of variation in power due to numerical artefacts. From the APS it is seen that Gaussian clustering of large numbers of very faint sources can produce anisotropy excesses of many orders of magnitude over the unclustered case. Increasing the number of Gaussian clusters decreases the anisotropy power as expected from Fig.~\ref{fig:ncluster_ps}. Figure~\ref{fig:condon_ps} demonstrates that Gaussian clustering of a very large number of faint sources can replicate the observed anisotropy power in the case where the clusters are small, $\lesssim 30''$.

\subsection{Anisotropy power from unsubtracted sources}
\label{sec:unsubtracted}

The measured APS is expected to have a contribution to the anisotropy power from known sources which were not removed by the subtraction process. As a first estimate, this contribution can be estimated by using a Franzen model with a cutoff in flux density to reflect that sources above a certain threshold have been subtracted. During data reduction, the auto-masking ensured all point sources greater than $5 \sigma$ or $\sim5$~mJy are subtracted (see Table~\ref{tab:field_details} for exact values). Therefore, a Franzen model with a cutoff at $5$~mJy could be used as one measure of the expected contribution to the APS from unsubtracted sources.

\begin{figure}
	\includegraphics[width=\columnwidth]{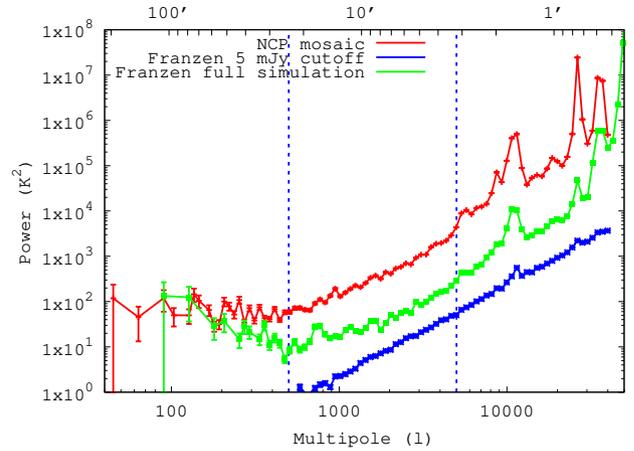}
    \caption{APS of the observed NCP mosaic and 2 methods of predicting the expected APS. The unclustered point source Franzen model with a simple flux density cutoff at 5~mJy is shown in blue as one prediction of the expected APS from unsubtracted sources. The APS of a more realistic simulation, taking the full unclustered point source Franzen model and performing cleaning and source subtraction with artificial noise present, is shown in green as a better prediction of the expected APS. The observed APS shows a significant excess over both cases.}
    \label{fig:excess_ps}
\end{figure}

However, selecting a realistic brightness upper limit is difficult, because this method does not account for artefacts or systematic errors introduced in the deconvolution and subtraction process. These could be manifested as areas of over-subtraction, calibration errors, or bright sources are not subtracted completely, although this contribution should be small. To overcome this difficulty, we perform a more complex but more realistic simulation. We take the whole Franzen model with no brightness cutoff, use {\sc{wsclean}} to simulate visibilities for the full Franzen model, and add artificial noise to the visibilities that is representative of the actual noise in the observations. The resulting visibilities are then run through the whole imaging process including deconvolution and subtraction. The resulting APS from this image is a more realistic predictor of the expected anisotropy power. The subtraction of sources is likely to be worse for the real data due to calibration artefacts and the ionosphere. A full simulation that includes all ionospheric, instrumental and calibration effects is considerably more complex. Therefore, the degree which the subtraction would be affected by these effects is deferred for future work. The APS for both the Franzen 5~mJy cutoff model and the cleaned Franzen model are presented alongside the observed APS of the mosaicked image in Fig.~\ref{fig:excess_ps}.

The observed power spectrum shows excess power at all scales over both methods of simulating the residual power. For large scales, the observed APS flattens off and becomes dominated by diffuse Galactic emission as expected. This should not occurs in the models as no Galactic emission was added. This is the case in the Franzen model with a simple flux density cutoff at 5~mJy. However, when the full model is taken as a starting point and cleaning and source subtraction with artificial noise present is performed, a deviation from the power law shape of the APS is seen for $\ell \lesssim 1000$. This feature is not expected for the unclustered point source model. At present this is unexplained and is likely due to an unknown artefact introduced during the subtraction process. Furthermore, this feature is not seen in the observed APS. However, for $\ell \gtrsim 1000$, corresponding to the apparent point source dominated region of the observed APS, the shape of the modelled APS is as expected. This is the region of interest for further analysis. An excess anisotropy power is observed here. This implies that unsubtracted sources expected from current source models are insufficient to explain the observed anisotropy of the radio background. In this region the excess appears to have a power law or multiplicative form. Henceforth, when the 'anisotropy excess' refers to the multiplicative anisotropy excess unless otherwise stated. However, it is possible that a constant anisotropy excess power also exists. This would be possible to observe as an excess over what is expected from Galactic synchrotron emission emission at low $\ell$. However, as this component is not included in our simulations, it is not possible to constrain whether an additive excess exists. Figure~\ref{fig:excess_ps} also shows that there are some artefacts, introduced during the cleaning and source subtraction process, which affect the APS.

\begin{figure}
	\includegraphics[width=\columnwidth]{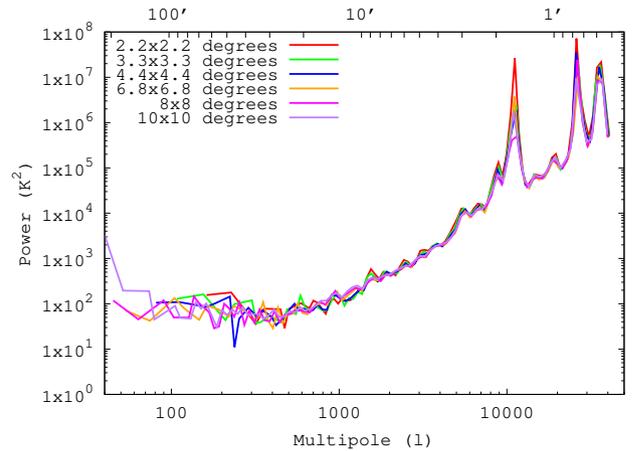}
    \caption{APS of the observed NCP mosaic for different sizes of image used to calculate the APS.}
    \label{fig:regions_ps}
\end{figure}

In order to check that the excess is not due to artefacts from the brightest sources, the APS can be calculated for different regions of the image. In the simplest case the APS can be calculated for smaller and smaller sized regions centred on the image centre. Since the brightest sources are rare and the image is not centred on a bright source, it would be expected that if residuals or artefacts from bright sources were contributing to the measured anisotropy excess then it would be expected that smaller regions would have less anisotropy power. The result of this is shown in Fig.~\ref{fig:regions_ps}. It can be seen that there is little change in anisotropy power between region sizes, implying that artefacts from individual bright sources are not the cause of the anisotropy power excess. However, this does not rule out the presence of some widespread effect, that affects both bright and faint sources.

The simulations in Fig.~\ref{fig:excess_ps} do not include large-scale structure (clustering of sources), diffuse Galactic emission or resolved sources. From measurements of large scale structure it is known that matter in the Universe is clustered \citep{chabanier_2019}. Galaxies and therefore radio sources trace this clustering. However, it is unlikely that large scale clustering can explain the observed excess, because the observed galaxy power spectrum has a form $C_l \propto \ell^{-1.2}$ as measured from NVSS data by \cite{blake_2004}. Where $C_{\ell}$ is the multipole moment of the radio galaxy power spectrum. Large-scale clustering with this galaxy power spectrum would manifest itself as a deviation from anisotropy power due to apparently unclustered sources. This is not seen in the shape of the measured APS. Large scale clustering is not observed in the APS as the brightest sources dominate the power as demonstrated in previous sections. Then, since the brightest sources are the least common they are least effected by statistical large scale clustering and this results in little change to the APS. The simulations do not include diffuse Galactic emission, however, the shape of the APS produced by this emission is flat \citep{gehlot_2022}. Therefore, Galactic diffuse emission cannot be the source of the power law anisotropy power excess, as it produces negligible anisotropy power at high $\ell$. Finally, all sources in the Franzen model are treated as point sources during the modelling. This is a valid assumption as around 90\% of radio sources are unresolved at 25 arcseconds resolution, the approximate resolution of our dirty beam while cleaning, as shown by observations similar to those in this work by \cite{procopio_2017}. It is unlikely that incorrectly modelling 10\% of sources as point sources instead of resolved could produce an excess anisotropy power on the level of that observed, especially since a multiscale clean is used for the observations in order to accurately model resolved sources for accurate subtraction. However, the idea that resolved sources can contribute to the anisotropy power excess is explored further in the next paragraph. Therefore, despite the simulations not including effects of large-scale structure, diffuse Galactic emission, or resolved sources, none of these would affect the modelled APS significantly enough to explain the observed anisotropy power law excess. However, a topic of future work will be to show this explicitly using similar modelling techniques.

\begin{figure}
	\includegraphics[width=\columnwidth]{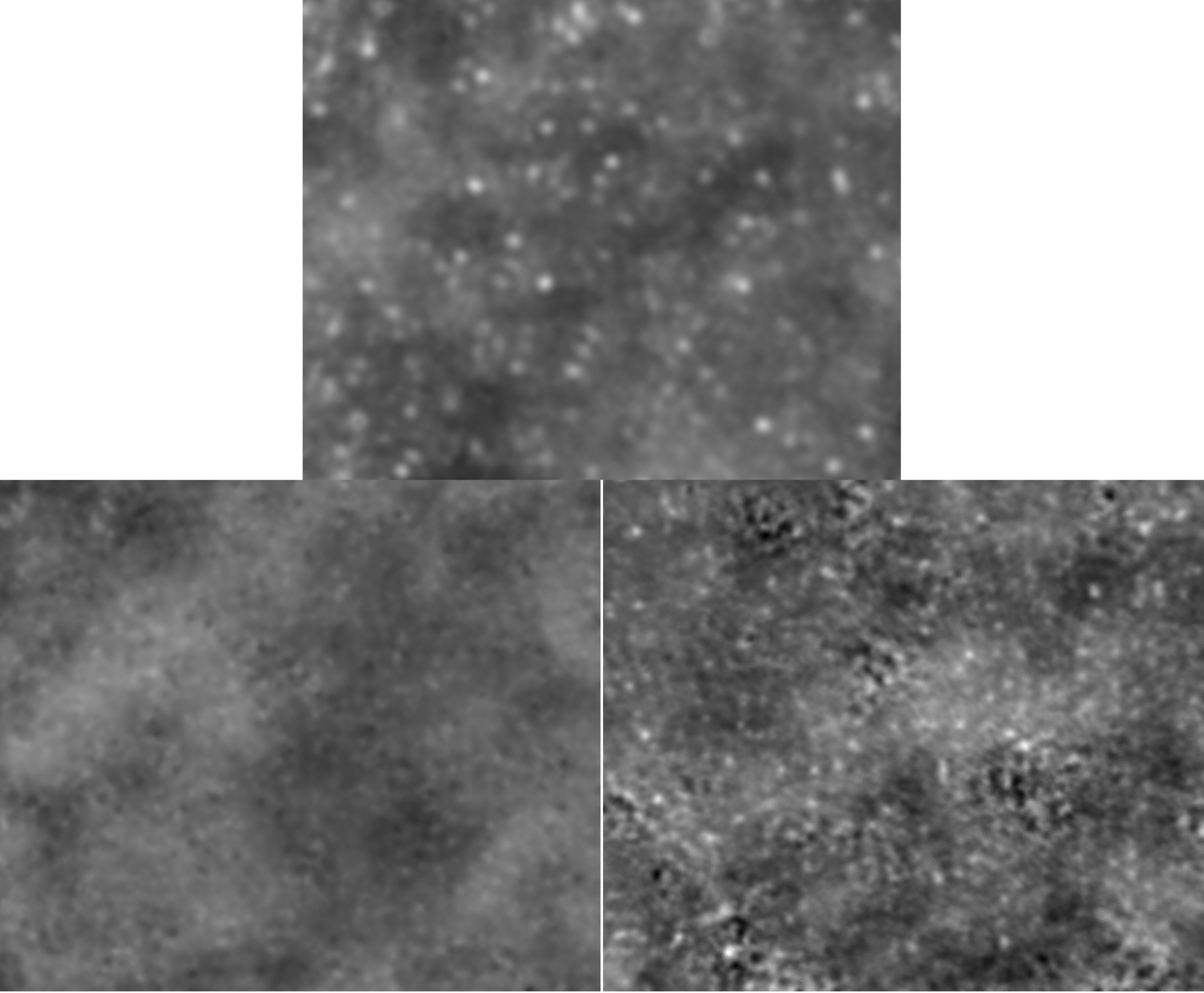}
    \caption{Top: a portion of the simulated image of the RSB from the unclustered Gaussian sources Franzen model, with cleaning and noise. In this model 50\% of the point sources are replaced with Gaussian sources of size 300 arcseconds. Bottom left: a portion of the simulated image of the RSB from the unclustered point source Franzen model simulation, with cleaning and noise. Bottom right: a portion of the observed background mosaic. All images are matched in flux density per beam scale and synthesised beam size. The image portions are of size $4.5^{\circ} \times 4.5^{\circ}$ and are naturally weighted.}
    \label{fig:fullsim_comp}
\end{figure}

Figure~\ref{fig:fullsim_comp} shows in the bottom right hand a portion of the observed background image and in the bottom left the simulated image of the background from the full Franzen model simulation with cleaning and noise. Both images are matched in brightness scale and are of size $4.5^{\circ} \times 4.5^{\circ}$. They are presented in natural weighting to emphasise diffuse structure. The key difference observed between the images is the presence of point like residuals in the real image. Most of these point like residuals do not seem to be associated with cleaned components, ruling out that they are artefacts from the subtraction or leftover from bright sources that have not been cleaned properly. However, these residuals also do not appear to have bright counterparts in the dirty uniformly-weighted image. This suggests that these are real resolved sources which have a peak flux density too low to be picked up by the $5\sigma$ auto-masking threshold during the clean. In naturally weighted images, which emphasise large-scale structure, resolved sources have a higher peak flux density and hence become visible. 

This idea was tested further by modifying our simulation from the unclustered Franzen model with noise and cleaning as described above. However, this time a certain proportion of the point sources in the model are replaced by extended smooth Gaussian sources of varying angular size. Fig.~\ref{fig:fullsim_comp} also shows the comparison between the point source only Franzen full simulation in the bottom left, and a Franzen full simulation where 50\% of the point sources were replaced by Gaussians of size 300 arcseconds in the centre top. This confirms that resolved sources with a lower peak flux density are not fully cleaned, despite using a similar auto-masking threshold. Additionally, the images of the observed RSB (Fig.~\ref{fig:fullsim_comp} bottom right) and the simulated diffuse sources (Fig.~\ref{fig:fullsim_comp} centre top) are similar, implying that the observed image is dominated by resolved sources.

\begin{figure}
	\includegraphics[width=\columnwidth]{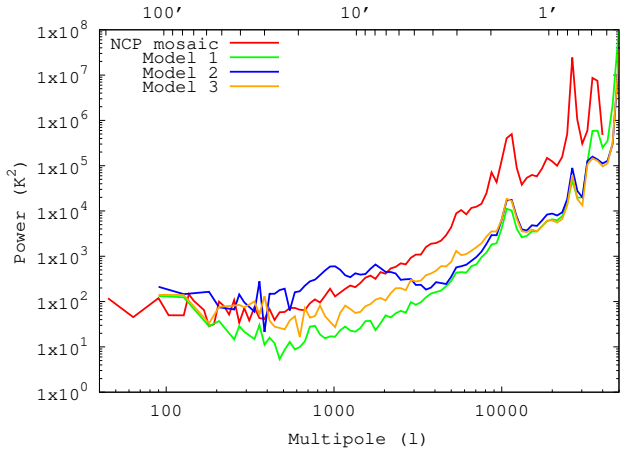}
    \caption{APS of unclustered Franzen models with cleaning and noise. Model 1 is a Franzen model with all sources rendered as point sources. Model 2 and 3 are Franzen models where 50\% of the point sources were replaced by Gaussian sources of sizes 300 and 50 arcseconds respectively. All models are cleaned to the same threshold and have the same noise level.}
    \label{fig:fullsim_ps}
\end{figure}

The fact that a population of diffuse, lower peak brightness sources is not removed by the cleaning presents one explanation for the observed anisotropy excess. These unremoved sources will contribute to the anisotropy excess as shown in Fig.~\ref{fig:cutoff_ps} and Fig.~\ref{fig:sourcesize_ps}. Additionally, since these diffuse sources are now the brightest sources left after subtraction, they likely dominate the anisotropy power. This means that the measured anisotropy power would be approximately equal to the contribution from these sources, as demonstrated by Fig.~\ref{fig:twopop_ps}. In order to directly test whether these unremoved diffuse sources are sufficient to explain the observed anisotropy excess, the APS was calculated for each of the full Franzen simulations with differing ratios of Gaussian sources. These simulations are not realistic but serve as a starting point for assigning fluxes to this population of diffuse sources. Fig.~\ref{fig:fullsim_ps} shows the APS for the full Franzen model simulation for models 1 to 3, where model 1 is only point sources, while for models 2 and 3, we replace half of the sources by Gaussian sources of size 300 and 50 arcseconds respectively. The presence of diffuse sources increases the anisotropy power of the background after source subtraction in all cases. The shape of the APS for the models where the diffuse source size is large shows that the excess disappears on scales roughly smaller than the source size. This is consistent with what is expected for anisotropy contributions for smooth diffuse sources as shown in Fig.~\ref{fig:sourcesize_ps}.

\begin{figure}
	\includegraphics[width=\columnwidth]{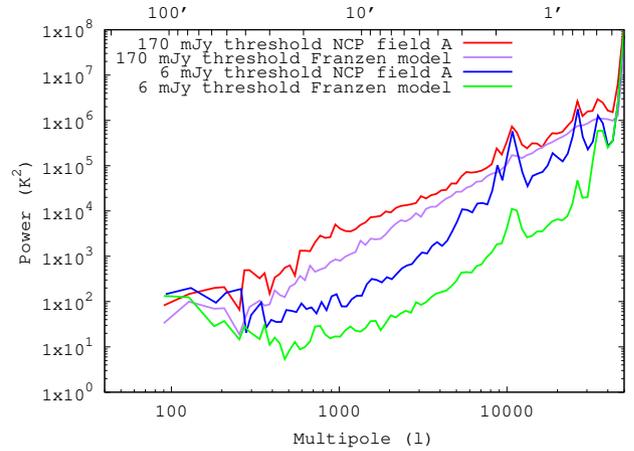}
    \caption{Observed APS of NCP field A with different auto-masking thresholds alongside full Franzen model simulations of the power expected from unsubtracted point sources, using the same respective subtraction thresholds.}
    \label{fig:thresh_ps}
\end{figure}

The observed excess can be probed to learn more about its properties. One example of this is investigating how the excess over our full Franzen model simulations changes depending on the auto-masking threshold used during cleaning. A higher auto-masking threshold leaves brighter sources in the image that would otherwise have been subtracted. Fig.~\ref{fig:thresh_ps} shows the observed power spectrum of NCP field A for two cases, where the auto-masking threshold was set to approximately 5~mJy and 170~mJy. The Franzen full simulations including artificial noise and source subtraction are included with the same respective auto-masking thresholds. This allows the excess to be studied at two different subtraction thresholds. From Fig.~\ref{fig:thresh_ps} it is seen that the observed APS depends strongly on the choice of subtraction threshold. The APS with the higher subtraction threshold shows an excess anisotropy over that with the lower threshold. Additionally, the higher subtraction threshold APS appears power law like over the whole scale probed, showing no signs of flattening at low $\ell$. This can be interpreted as the unclustered point source contribution to the APS dominating over the contribution from diffuse Galactic emission. Furthermore, it is seen that the excess over the expected anisotropy power is multiplicatively larger at lower subtraction thresholds. This is expected if the unsubtracted Franzen sources begin to dominate the APS at higher subtraction thresholds. However, the excess is additively slightly larger at higher thresholds, although this is not apparent due to the logarithmic scale of the graph. This may suggest that the cause of the excess is affected during the source subtraction process. Assuming the cause of the excess to be some unknown population of sources, possible explanations for this include subtracting this population around known sources during subtraction, whether these sources are correlated with known sources or not, or beginning to directly subtract these sources if they are bright enough to be picked up during cleaning. A more complete analysis of how the anisotropy power at a particular scale varies with subtraction threshold may be able to allow constraints to be placed on the differential source count of unknown sources, using equation~\ref{unclustered_power}. This is deferred for future work.

It is useful to obtain an equation of the measured anisotropy power of the RSB at 120 MHz for comparison to past and future measurements. In the range $50\leq \ell \leq 4000$ the APS (in squared units) is well fit by a power law plus constant of the form:
\begin{equation}
\label{eq:APS}
    (\Delta T)_{\ell, \nu}^2 = ((3 \pm 2)\times 10^{-5}\ell^{2.17\pm0.08} + (41\pm7))\nu^{-5.32} \,\text{K}^2.
\end{equation}
Here we assume that the anisotropy power of the RSB scales in frequency like the RSB itself (see equation~\eqref{eq:background}), but this is squared due to the $\text{K}^2$ units of anisotropy power. This may not be a valid assumption as the scaling of of the APS with frequency may not be trivial. Additionally, this scaling may be a function of $\ell$. The above equation is only valid for subtraction thresholds close to 5~mJy. The scaling with subtraction threshold is not trivial and depends on the differential source count and clustering properties of the radio source populations. For the case of clustered sources the scaling may also be a function $\ell$. Additionally, it is useful to have an equation for the multiplicative anisotropy power excess between the observed APS of the RSB and the expected anisotropy power from unsubtracted point sources. Taking the difference between the observed APS and the Franzen full simulation APS in Fig.~\ref{fig:excess_ps}, a power law is  fit to the excess between an $\ell$ of 700 and 4000 has the form: 
\begin{equation}
\label{eq:excess}
    (\Delta T)_{\ell, \nu}^2 = ((7 \pm 6)\times 10^{-4}\ell^{1.7\pm0.11})\nu^{-5.32} \,\text{K}^2,
\end{equation}
where similar assumptions to equation~\eqref{eq:APS} have been made.

\section{Discussion}
\label{sec:disscussion}

Our APS of the RSB, measured using LOFAR, can be compared to other measurements of the APS, both with LOFAR and other instruments. Our measured angular power for scales not dominated by diffuse Galactic structure is similar to that reported in \cite{choudhuri_2020} for the 4 fields presented in their Fig.~1, after conversion from the $C_{\ell}$ to the $(\Delta T)^2_{\ell}$ normalisation (see appendix A in \citealt{offringa_2022}). A more robust comparison is difficult as for each field the subtraction threshold is not explicitly reported, and as shown in Fig.~\ref{fig:thresh_ps}, this is a crucial parameter that strongly influences the power spectrum at high $\ell$. The measured angular power depends heavily on the subtraction threshold used as it is the brightest remaining sources which will dominate the power as shown in \S \ref{sec:results}. Note \cite{choudhuri_2020} similarly report an excess anisotropy power (of approximately 2 orders of magnitude) over their model of sources up to 50~mJy. This model also included large scale clustering of radio sources using the angular correlation function from \cite{dolfi_2019}. Their model is in good agreement with our models of unclustered point sources for $\ell>1000$. Therefore, as justified in \S \ref{sec:unsubtracted}, large scale clustering appears to make a negligible difference to the APS at small scales.

In the angular scales of overlap, in this case the largest scales probed by our measurements, our observations of the APS are in excellent agreement with results presented in \cite{gehlot_2022}. Their measurements are made  also centred on the NCP and at the same frequency, but with an effectively independent instrument in terms of systematic, resolution and $uv$-coverage. This is further indication that our observations are well calibrated and do not suffer from any major systematic errors.

\begin{figure}
	\includegraphics[width=\columnwidth]{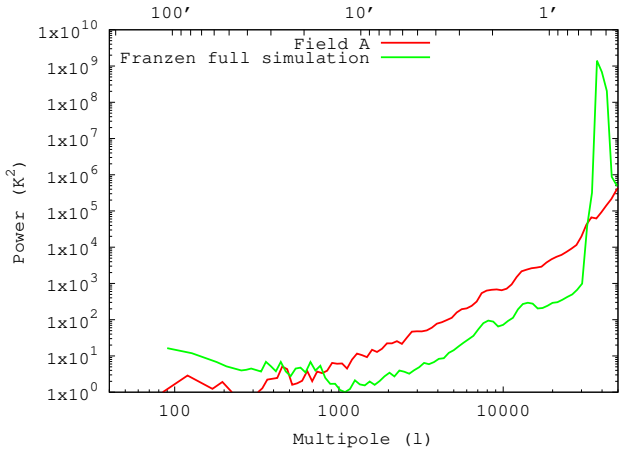}
    \caption{The APS of the field A observed by \protect\cite{offringa_2022} at 140~MHz. Also presented is the expected anisotropy power from unsubtracted point sources from a full Franzen simulation as detailed in Section~\ref{sec:unsubtracted}.}
    \label{fig:fieldA_ps}
\end{figure}

Our previous results, reported in \cite{offringa_2022}, are shown in Fig.~\ref{fig:fieldA_ps} alongside a full Franzen model simulation of the expected APS, with representative artificial noise added, using the method detailed in Section~\ref{sec:unsubtracted}. Figure~\ref{fig:fieldA_ps} shows that with the newer and more accurate simulation of the expected APS, an excess anisotropy power is still observed in the previous measurement.  Our measured angular power is (in squared units) approximately 25 times above these previous results reported in \cite{offringa_2022}. However, this is expected, as the source subtraction threshold there was around 3.3 times lower than in this work due to a lower noise in the observed fields. The true scaling with subtraction threshold depends on the populations contributing to the anisotropy power and so is not trivial. Assuming the anisotropy power is only due to sources in the Franzen model, integrating equation~\eqref{unclustered_power} using \eqref{franzen} can be used to show that the scaling factor due to differing subtraction thresholds is approximately 5.7. This, combined with the expected frequency scaling for the RSB given in equation \eqref{eq:excess} (which comes to a factor of approximately 2.2) brings the previous results to within a factor of 2 with those presented in this work. In \cite{offringa_2022} an apparent scaling of the anisotropy power was observed with the square of the average brightness temperature of the different fields, calculated using the radio sky map presented in \cite{haslam_1982}. If this scaling is used then an additional factor of 3.4 is applied to the APS result for Field A relative to \cite{offringa_2022} for a comparison to the measurements in this work. This is because the NCP has a lower average brightness temperature than Field A, according to the radio sky map. If the difference in brightness temperature between the fields is due only to diffuse Galactic emission, we see no reason to scale the APS by this factor. This is because diffuse Galactic emission has a flat APS, and so at high $\ell$ it is negligible compared to the point source contribution to the anisotropy power. This means any increase in the diffuse Galactic emission brightness should not scale the APS at high $\ell$. Instead we propose that the factor of two discrepancy between this work and the previous is due to the non-trivial subtraction threshold or frequency scaling.

The results presented in this work suggest it is likely that measured excess anisotropy power over what is predicted by semi-empirical models for point sources is of astrophysical origin and not due to systematic errors. Evidence for this is presented in Fig.~\ref{fig:regions_ps}, demonstrating there are no significant artefacts associated with subtraction of bright sources. Additionally, inspecting the mosaicked images no bright artefacts are found. Furthermore, fainter artefacts associated with all sources are also unable to cause the excess. This is because faint artefacts are unlikely to be able to dominate the anisotropy power as would be required to produce the large observed excess, as demonstrated in Fig.~\ref{fig:cutoff_ps}. Finally, the flux calibration presented in \S\ref{sec:obs} shows that the pre-processing of the data is accurate and no large flux calibration errors exist. However, the possibility of the excess arising from calibration errors is not completely ruled out.

The APS of the RSB measured in this work and our previous work can be used to constrain the possible causes for the RSB surface brightness excess level above that expected from known classes of radio sources. It is possible that the observed anisotropy excess of the RSB and the observed surface brightness excess of the RSB have different origins. For example, the observed  surface brightness excess could be due to a smooth contribution to the RSB and the observed anisotropy excess due to calibration errors. However, the existence of some form of anisotropy excess over current models is expected if there is a surface brightness excess and if the cause of the surface brightness excess is not completely smoothly isotropic.

If the multiplicative anisotropy excess and surface brightness excess are of common origin then the shape of the anisotropy excess can rule out certain causes of the observed surface brightness excess. The anisotropy excess appears constant on a logarithmic scale and so is of a power law form. This makes it unlikely that large scale diffuse emission from the Milky Way is the cause, as this has a constant valued or flat APS \citep{gehlot_2022}. For the same reasons, large scale smooth extragalactic emission is also an unlikely cause.

Instead the simple power-law shape of the multiplicative anisotropy excess resembles that which would be caused by a population of point-like, unclustered sources. However, as discussed in \S \ref{sec:unsubtracted} and \citealt{offringa_2022} we believe that known classes of sources cannot contribute this level of anisotropy power.  It is possible that the anisotropy and surface brightness excesses are caused by a new population of point sources that have not been detected. However, these point sources would have to be faint to have remained undetected by source counts and extremely numerous in order to create the surface brightness excess. If these faint point sources were unclustered then as seen from Fig.~\ref{fig:cutoff_ps} and Fig.~\ref{fig:twopop_ps} they are unable to create any observable anisotropy excess over current models. However, clustering of these sources can create more anisotropy power, potentially explaining the excess. As shown in Fig.~\ref{fig:condon_ps}, clustering of many faint sources can create massive excesses in anisotropy power, and in general this could explain the observed excess. In order for this to be the case the clustering must be extreme, in the sense that each cluster contains many sources. Additionally, the cluster size must be on the order of a few arcminutes or smaller. This is because, as seen in Fig.~\ref{fig:clustersize_ps}, on scales smaller than the cluster size, the expected APS shape deviates from unclustered point-source like. This is not seen in the observed APS, at least up to $\ell=4000$, where instrumental power spectrum systematics begin to dominate. Extreme clusters such as these produce varied observational signatures in images. However, for cases with a large number of small clusters, such as the Condon model with 100,000 clusters which is shown in Fig.~\ref{fig:condon_ps}, the associated naturally weighted images were observed to show a relatively smooth background, with large scale variations of order 10~mJy. Furthermore, the associated uniformly weighted images appear noise-like with a small number of peaks around 1~mJy. Neither of these images are inconsistent with our observations, and it is possible that a clustered point source population has gone undetected in deep source counts to date.

Another possible explanation for the observed anisotropy excess which would also cause an associated surface brightness excess, is the existence of a new population of resolved diffuse sources. After inspection of the naturally weighted mosaicked image of the background it can be seen that there are point source-like residuals apparently above the cleaning threshold. These sources have no bright counterpart in the uniform weighted image. This suggests that these sources are diffuse, accentuated by the natural weighting of the image. As demonstrated in Fig.~\ref{fig:fullsim_comp} these diffuse sources are not picked up by the cleaning in uniform weighting due to their low peak flux density, an effect which is present in source count measurements and known as resolution bias \citep{mandal_2021}. As a result of this, as shown in Fig.~\ref{fig:fullsim_ps}, a population of diffuse sources have the potential to create the measured anisotropy excess after source subtraction. However, the APS of resolved sources decays at large $\ell$ corresponding to scales smaller than the resolved source size, as seen in Fig.~\ref{fig:sourcesize_ps} and Fig.~\ref{fig:fullsim_ps}. This decay is not visible in the measured APS, and therefore if resolved sources are the cause of the multiplicative anisotropy excess they likely have a size of order 1 arcminute or smaller. Current low-frequency deep source counts have not well probed this parameter space of faint diffuse sources, for example sources of size ~1 arcminute and flux less than 10~mJy \citep{mandal_2021}. New diffuse sources have been previously investigated and ruled out as a possible cause of the RSB  surface brightness excess using confusion analysis by \cite{vernstrom_2015}, but only at higher frequencies and for sources up to 2 arcminutes and with total flux density greater than $\sim$1~mJy. Therefore, it is possible that diffuse sources of several arcminutes are responsible for the observed anisotropy and surface brightness excesses, and a qualitative analysis of our images provides support to this hypothesis. 

\section{Conclusion}
\label{sec:conclusion}

We have performed measurements with LOFAR to determine the anisotropy angular power of the radio background at 120~MHz on angular scales from $3^\circ$ to $0.3^\prime$. As discussed in \S\ref{sec:obs} our data comes from 12 hours of observations of 7 fields in the vicinity of the NCP. As discussed in \S\ref{sec:modelling} we have performed detailed simulations of the radio sky and the associated anisotropy power based on two source count models in the literature, to better understand the effect of different variables on the measured APS. As shown in \S\ref{sec:results} we find contributions to the APS of the RSB from both diffuse Galactic structure and apparently unclustered point sources. We also find that many variables of the source population can effect the measured APS of the background. As demonstrated in Fig.~\ref{fig:excess_ps} we find that our measured angular power on almost all scales is in significant excess of what could be caused by known point source populations.

As discussed in \S\ref{sec:disscussion} the results presented in this work suggest that a potentially promising cause of the measured multiplicative anisotropy excess in the RSB is a population of diffuse sources, currently unaccounted for in source counts {\bf because the sources remain undetected in those surveys}. This would also provide an explanation for the excess surface brightness of the RSB. Theoretical models exist which predict diffuse sources, for example, dark matter decay or annihilation \citep[e.g.,][]{fornengo_2011,hooper_2012} or cluster mergers \citep[e.g.,][]{fang_2016}. A population of very faint point sources with extreme clustering is also a possible explanation for the multiplicative anisotropy excess and the  surface brightness excess. However, both of these explanations predict a deviation from the point source like shape of the APS at large $\ell$, which is not observed. This places restrictions on the maximum size of the resolved sources and clusters if they are responsible for the multiplicative anisotropy excess of the RSB.

Future work will consist of performing more full pipeline simulations, including new populations of diffuse and clustered sources alongside known sources, in order to understand what the flux distribution and number of the population could be. Moreover, further observations and analyses are planned to investigate the variation of the APS with frequency and subtraction threshold, in order to better understand the cause of the excess. Additionally, the full pipeline simulation will be improved to account for more systematic uncertainties such as calibration errors. Furthermore, more realistic full cosmological modelling will be done to find what distributions of point sources could potentially produce the observed APS. This will demonstrate what 3-dimensional clustering of point sources is need to replicate observations and whether this is physically realistic. Finally, a cross-correlation analysis of images of the RSB with different measures of large scale structure in the universe such as galaxy catalogues and CMB lensing surveys is needed. This will enable the search for a correlation signal between the source population contributing to the anisotropy of the RSB and large scale structure, to understand the nature of this potentially undetected source population.

\section*{Acknowledgements}

We thank the ASTRON/JIVE Summer Research Program for supporting F. J. Cowie during the course of the research. 
S. Heston is supported by NSF Grant No. PHY-1914409 and the U.S.~Department of Energy Office of Science under award number DE-SC0020262. 
The work of SH is supported by the U.S.~Department of Energy Office of Science under award number DE-SC0020262, NSF Grant No.~AST1908960 and No.~PHY-1914409 and No.~PHY-2209420, and JSPS KAKENHI Grant Number JP22K03630. This work was supported by World Premier International Research Center Initiative (WPI Initiative), MEXT, Japan.

\section*{Data Availability}

The data underlying this article will be shared on reasonable request to the corresponding author.



\bibliographystyle{mnras}
\bibliography{version_1} 





\bsp	
\label{lastpage}
\end{document}